\documentstyle[11pt,amssymbols,epsf]{article}
\renewcommand{\l}{\newline\null}

\newcommand{\be}{\begin{equation}}
\newcommand{\ee}{\end{equation}}

\newcommand{\ba}{\begin{array}}
\newcommand{\ea}{\end{array}}

\newcommand{\bea}{\begin{eqnarray}}
\newcommand{\eea}{\end{eqnarray}}

\newcommand{\rar}{\rightarrow}
\newcommand{\p}{\partial}
\newcommand{\ol}{\overline}

\newcommand{\la}{\langle}
\newcommand{\ra}{\rangle}
\def\figskip{\vskip .5cm plus 3mm minus 2mm}
\begin{document}
\begin{titlepage}
May 1993\hfill PAR-LPTHE 93/18
\begin{flushright} hep-ph/9305212 \end{flushright}

\vskip 5cm
\begin{center}
{\bf THE STANDARD MODEL OF LEPTONS AS A PURELY VECTORIAL THEORY}
\end{center}
\centerline{\sl {(revised and augmented version)}}
\vskip 1cm
\centerline{M. Bellon
         \footnote[1]{Member of Centre National de la Recherche Scientifique}
        \& B. Machet \footnotemark[1]}
\vskip 5mm
\centerline{{\em Laboratoire de Physique Th\'eorique et Hautes Energies}
     \footnote[2]{LPTHE tour 16/1er \'etage,
      Universit\'e P. et M. Curie, BP 126, 4 place Jussieu,
          F 75252 PARIS CEDEX 05 (France).}
      \footnote[3]{E-mail: bellon@lpthe.jussieu.fr {\em or}
                           machet@lpthe.jussieu.fr},
                                                          {\em Paris}}
\centerline{\em Unit\'e associ\'ee au CNRS URA 280.}
\vskip 2cm
{\bf Abstract.}  We propose a way to reconcile the Standard Model of leptons
with a purely vectorial theory. The observed neutrino is predicted to
be massless. The unobservability of its partner and the $V-A$ structure
of the weak currents are given the same origin.
\smallskip

\vfill
\null\hfil\epsffile{LOGO.eps}
\end{titlepage}

\section {Introduction}

The Standard Model \cite{GlashowSalamWeinberg} is the presently accepted
description of the leptonic interactions. Its most
puzzling aspects stay anyhow the $V-A$
structure of the weak currents and the experimental absence of a right-handed
neutrino. We connect here both phenomena: the structure of the weak currents is
related to the unobservability of one (Majorana) component of the neutrino
which
gets an infinite mass; the other component, corresponding to the observed
neutrino, is strictly massless.  This arises in a purely vectorial theory
endowed with a composite scalar multiplet.

We describe here an anomaly-free leptonic sector.
We can forget about the quarks since we make the hadronic
sector anomaly-free too in refs.\cite{Machet,BellonMachet1}; in the same works
the gauge boson mass generation is presented as a hadronic phenomenon.

\section{The leptonic Lagrangian}

We only consider here 1 family of leptons, and we will not question the
$e-\mu-\tau$ universality.

Let $\Omega$ be the leptonic doublet
\be
\Omega = \Biggl ( \ba{c}  \ell^-  \\
                         \nu
         \ea \Biggr ),
\ee
with hypercharge $Y= Q- T^3 = -1/2$.
The starting Lagrangian is
\be
{\cal L} =
i\, \ol\Omega\  \gamma^\mu (\p_\mu -ig'B_\mu{\Bbb Y} -ig \vec W_\mu.\vec{\Bbb
T})\  \Omega
- m \ \ol\Omega \Omega.
\ee
$\vec{\Bbb T} = \vec\tau /2$, with $\vec\tau$ the Pauli matrices.
The interaction being purely vectorial, the mass term is gauge invariant and
there is no anomaly.

We introduce the 2 Majorana neutrinos
\be \left \{ \ba {lcl}
                        \chi &=& \nu_L + (\nu_L)^c, \\
                        \omega &=& \nu_R + (\nu_R)^c.
    \ea \right .
\ee
The superscript ``c'' means ``charge-conjugate''.
In terms of $\chi$ and $\omega$, $\cal L$ can be written (we forget hereafter
the `$^-$' superscript for $\ell$)
\be \ba {lcl}
{\cal L} &=&
i\, \ol\ell\,\gamma^\mu\p_\mu\ell +{i\over 2}\, \ol\chi\gamma^\mu\p_\mu\chi +
{i\over 2}\,  \ol\omega\gamma^\mu\p_\mu\omega  \cr
& & {}+ \bigl (\;\ol\ell\; ,\; \ol\chi\; \bigr )\  \gamma^\mu(g'B_\mu{\Bbb Y}
+g \vec W_\mu.\vec{\Bbb T})
\ {1-\gamma_5\over 2}\ \Biggl ( \ba {c} \ell \\ \chi \ea \Biggr ) \cr
& & {}+ \big (\; \ol\ell\; ,\; \ol\omega\;\bigr )\  \gamma^\mu(g'B_\mu{\Bbb Y}
+g \vec W_\mu.\vec{\Bbb T})
\ {1+\gamma_5\over 2}\ \Biggl ( \ba {c} \ell \\ \omega \ea \Biggr) \cr
& & {}-{1\over 2} m\  (\ol\chi\omega + \ol\omega\chi + 2\ol\ell\ell).
\label{eq:L}\ea\ee

\section{Introducing a scalar triplet}

We introduce a scalar triplet $\Delta$ of composite fields,
with leptonic number $2$, according to
\be
\Delta = \left ( \ba {l} \Delta^{0} \cr
                         \Delta^{-} \cr
                         \Delta^{--} \cr
         \ea \right )
       ={\rho\over\nu^3}
          \left ( \ba {c} \ol{\omega_L} \omega_R \cr
                          {1\over \sqrt{2}}(\ol{\omega_L} \ell_R +
                                    \ol{(\ell_R)^c} \omega_R) \cr
                          \ol{(\ell_R)^c} \ell_R
          \ea \right )
       = {\rho\over\nu^3}
          \left ( \ba {c} \ol{\nu^c}\ {1+\gamma_5\over 2}\ \nu \cr
                       {1\over \sqrt{2}}(\ol{\ell^c}\ {1+\gamma_5\over 2}\ \nu
+
                                    \ol{\nu^c}\ {1+\gamma_5\over 2}\ \ell) \cr
                          \ol{\ell^c}\ {1+\gamma_5\over 2}\ \ell
          \ea \right ).
\ee
It is a triplet of $SU(2)$ (right). Its hermitian conjugate is
\be
\ol\Delta = \left ( \ba {l} {\bar\Delta^{0}} \cr
                            \Delta^{+} \cr
                            \Delta^{++} \cr
         \ea \right )
       ={\rho\over\nu^3}
          \left ( \ba {c} \ol{\omega_R} \omega_L \cr
                          {1\over \sqrt{2}}(\ol{\ell_R} \omega_L +
                                   \ol{\omega_R} (\ell_R)^c) \cr
                          \ol{\ell_R} (\ell_R)^c
          \ea \right )
       = {\rho\over\nu^3}
          \left ( \ba {c} \ol{\nu}\ {1-\gamma_5\over 2}\ \nu^c \cr
                      {1\over \sqrt{2}}(\ol{\ell}\ {1-\gamma_5\over 2}\ \nu^c +
                                    \ol{\nu}\ {1-\gamma_5\over 2}\ \ell^c) \cr
                          \ol{\ell}\ {1-\gamma_5\over 2}\ \ell^c
          \ea \right ).
\ee
We can impose the conditions
\be
\la\ol{\omega_L}\, \omega_R\ra = \la\ol{\omega_R}\, \omega_L\ra =
\nu^3 \not = 0,
\ee
or, equivalently
\be
\la \Delta^0\ra = \la\ol{\Delta ^0}\ra = \rho.
\ee
Indeed, as soon as the mass of $\omega$ is not identically $0$, the
electroweak fluctuations depicted in fig.\nobreak 1 have no reason to vanish.
\figskip
\epsffile{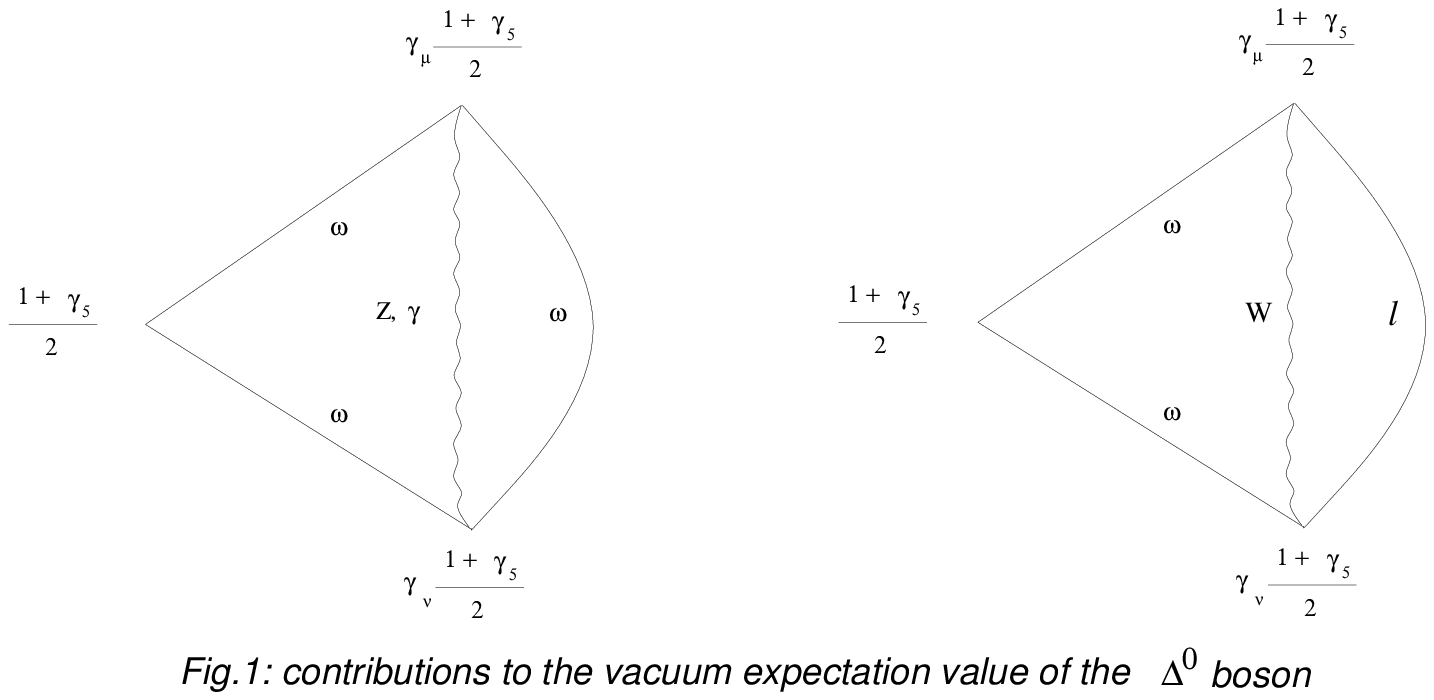}

\figskip
The scalars being made up with fermions, the path integrations over those
degrees of freedom cannot be performed independently; we thus
introduce (like in the hadronic sector
\cite{Machet,BellonMachet1,BellonMachet2})
constraints that we exponentiate into the effective Lagrangian
($\Lambda$ is an arbitrary mass scale)
\be\ba{lcl}
{\cal L}_c &=& \lim_{\beta\rar 0}
            -{\Lambda^2\over 2\beta}\left[
\left (\Delta^0 -{\rho\over\nu^3}\, \ol{\omega_L} \omega_R\right )
\left (\ol{\Delta^0}-{\rho\over\nu^3}\, \ol{\omega_R} \omega_L\right )
\right .\cr
& &+
\left (\Delta^- -{1\over\sqrt{2}}{\rho\over\nu^3}(\ol{\omega_L} \ell_R +
\ol{(\ell_R)^c} \omega_R) \right )
\left (\Delta^+ -{1\over\sqrt{2}}{\rho\over\nu^3}(\ol{\ell_R} \omega_L +
\ol{\omega_R} (\ell_R)^c) \right )\cr
& &+
\left.\left (\Delta^{--} -{\rho\over\nu^3}\, \ol{(\ell_R)^c} \ell_R\right )
\left (\Delta^{++} -{\rho\over\nu^3}\, \ol{\ell_R} (\ell_R)^c\right )\right ].
\label{eq:Lc}\ea\ee

\section{Effective 4-leptons couplings; the mass eigenstates}

The equations (\ref{eq:L}) and (\ref{eq:Lc}) yield a ``see-saw'' mechanism
\cite{GellMannRamondSlansky} in the neutrino sector. Indeed, when
$\la\Delta^0\ra = \rho$, ${\cal L}_c$ gives
 the $\omega$ neutrino an infinite bare Majorana mass,
\be
M_0 = -\frac {\Lambda^2 \rho^2}{\beta\nu^3},
\ee
the $\chi$ neutrino a vanishing (Majorana) mass, and a finite Dirac mass
connects $\chi$ and $\omega$.
We have to diagonalize the mass matrix to get the mass eigenstates,
(see for example \cite{ChengLi}); they are the Majorana neutrinos
$\chi$ and $\omega$ themselves, and correspond to mass eigenvalues
$0$ and $\infty$ respectively.  The charged lepton keeps its Dirac mass $m$.

However, 4-fermions couplings may alter the mass spectrum, together with
being an obstacle for renormalizability. We propose to build a reshuffled
perturbative expansion based not on the `bare' 4-fermions couplings occurring
in ${\cal L}_c$, but rather on effective couplings obtained by resumming
infinite series of `ladder' diagrams as proposed by Nambu and Jona-Lasinio
\cite{NambuJonaLasinio}. We however differ from them by
bare couplings and a bare fermion mass both infinite; this
makes the effective couplings vanish with $\beta$, and the ``see-saw''
mechanism above stay unaltered.

In all figures below, the $L$, $R$'s at the vertices stand for the
projectors $(1-\gamma_5)/2$ and $(1+\gamma_5)/2$.

\subsection{The $4\,\omega$ coupling and the $\omega$ mass}

Let $\zeta_{\omega\omega}(q^2)$ be the $4\,\omega$ effective coupling
defined by resumming the geometric series depicted in fig.~2:
\figskip
\hskip -1.5cm
\epsffile{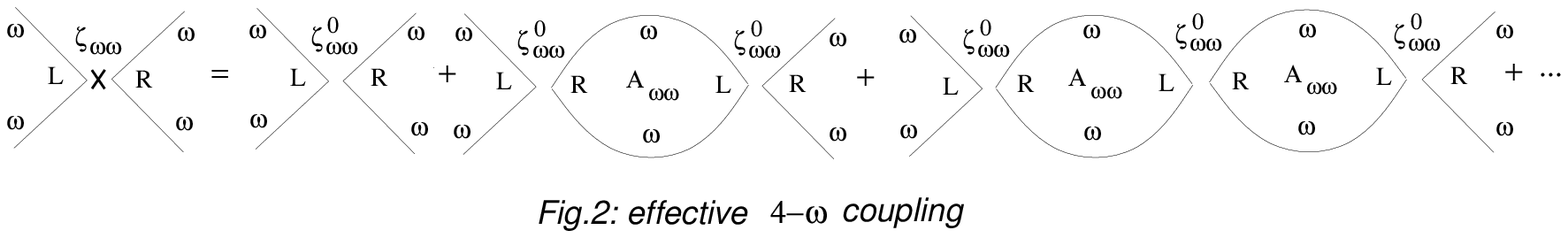}
\figskip
When $\la\ol\omega\omega\ra \not = 0$, $\zeta_{\omega\omega}$ contributes to
the $\omega$
mass according to fig.~3 (the $L,R$ projectors forbid fig.~4 with
$\ell\ell\omega\omega$ coupling to contribute).
\figskip
\hskip 3cm
\epsffile{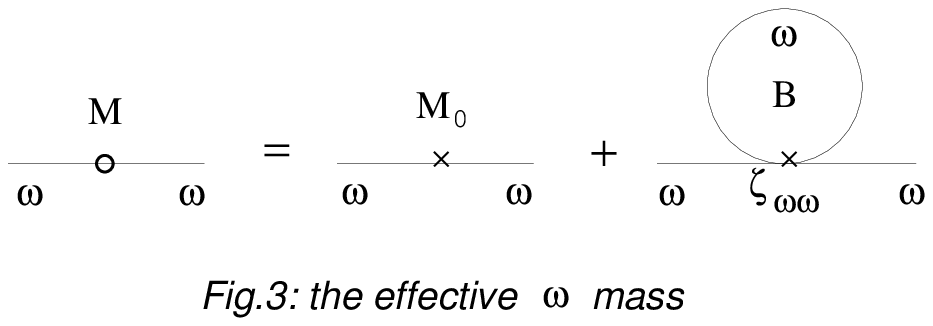}

\hskip 4cm
\epsffile{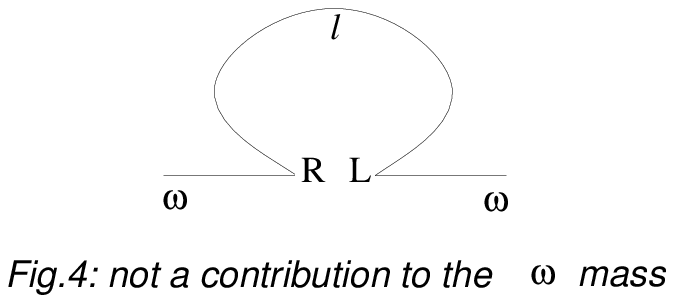}
\figskip
Figs.~2,3 yield the two coupled equations
\be
\zeta_{\omega\omega}(q^2)=
\frac {\zeta_{\omega\omega}^0}
{1-\zeta_{\omega\omega}^0 \, A_{\omega\omega}(q^2, M)}
\label{eq:zeta},
\ee
and
\be
M = M_0 - B\,  \zeta_{\omega\omega}(0)\label{eq:M},
\ee
where $\zeta_{\omega\omega}^0$ is the bare $4\,\omega$ coupling appearing in
${\cal L}_c$, behaving like $\beta^{-1}$
\be
\zeta_{\omega\omega}^0= -{\Lambda^2 \rho^2\over 2\beta \nu^6} =
{M_0\over 2\nu^3}.
\ee
$A_{\omega\omega}(q^2, M)$ is the 1-loop $\omega$ fermionic bubble, and
$B=\la\ol\omega\omega\ra =2\nu^3$. Notice that the effective coupling
{\em at 0 momentum transfer} is involved in (\ref{eq:M}).

$B$ being finite, $M=M_0$ is a solution of (\ref{eq:M}) as soon as
$\zeta_{\omega\omega}(0)$ goes to $0\;$; this is precisely the case as seen
from (\ref{eq:zeta}) since $A(0,M_0)$ involves a term $b\; M_0^2 +\cdots\;$
(see for example \cite{Broadhurst}). It corresponds to an effective
coupling
\be
\lim_{\beta\rar 0}^{q^2\rar\infty} \zeta_{\omega\omega}(q^2) =
-{1\over a \, q^2 +b M_0^2},
\ee
vanishing like $\beta^2$ when $\beta\rar 0$. ($a$ and $b$ in the formulas
above are numerical coefficients.)

\subsection{The ${\bf\ell\ell\omega\omega}$ coupling}

We define similarly the effective $(\ell\ell\omega\omega)$ coupling by the
series depicted in fig.~5. The same argumentation shows that it
behaves like $(-)A_{\ell \omega}(q^2,m,M_0)^{-1}$, where
$A_{\ell \omega}(q^2,m,M_0)$ is the fermionic loop involving one $\ell$ and
one $\omega$, and thus vanishes like $\beta^2$.
\figskip
\hskip -1.5cm
\epsffile{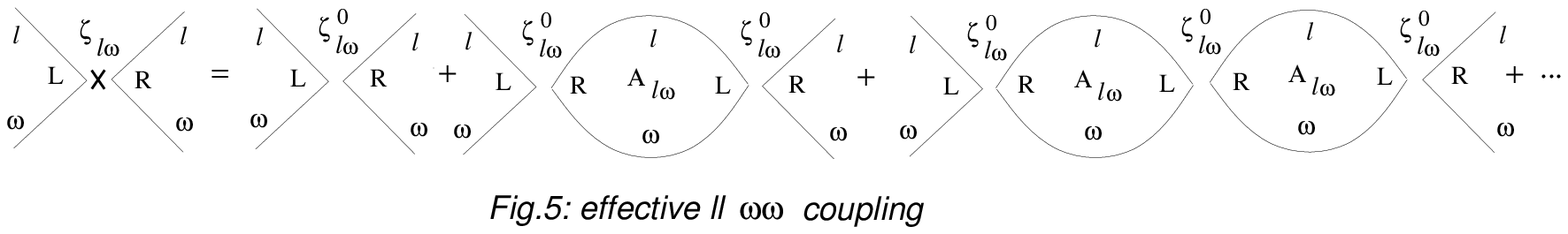}
\figskip

\subsection{The ${\bf 4\ell}$ coupling}

The situation here is more delicate since the same type of resummation as
above leads to an effective coupling behaving like a polynomial of degree
$0$ in $\beta$, the fermionic loop involving no infinitely massive particle.
However, if we go beyond the Nambu-Jona-Lasinio approximation, which
corresponds, in the language of \cite{NambuJonaLasinio} to propagating
$\Delta$ bound states (see figs.~2,5), we can show that
the presence of $\ell\ell\omega\omega$ couplings, by introducing $\omega$
loops, makes the bare $4\,\ell$ coupling $\zeta_{\ell\ell}^0$
 exactly cancel with the four series analogous to that of fig.~6, which writes
\be
\tilde\zeta_{\ell\ell}
= \zeta_{\ell\omega}^0\  \frac{A(\omega,\omega)}{1 -{A(\omega,\omega)}
     \;\zeta_{\omega\omega}^0}\; \zeta_{\ell\omega}^0
\approx -{\zeta_{\ell\omega}^{0\;2}\over \zeta_{\omega\omega}^0}
= -{1\over 4}\zeta_{\omega\omega}^0
= -{1\over 4}\zeta_{\ell\ell}^0.
\ee
The  $\zeta_{\ell\omega}^0$ and $\zeta_{\omega\omega}^0$ are the bare
$\ell\ell\omega\omega$ and $4\,\omega$ couplings of ${\cal L}_c$;
$A(\omega,\omega)$ is the fermionic $\omega$-loop involved here.
\figskip
\epsffile{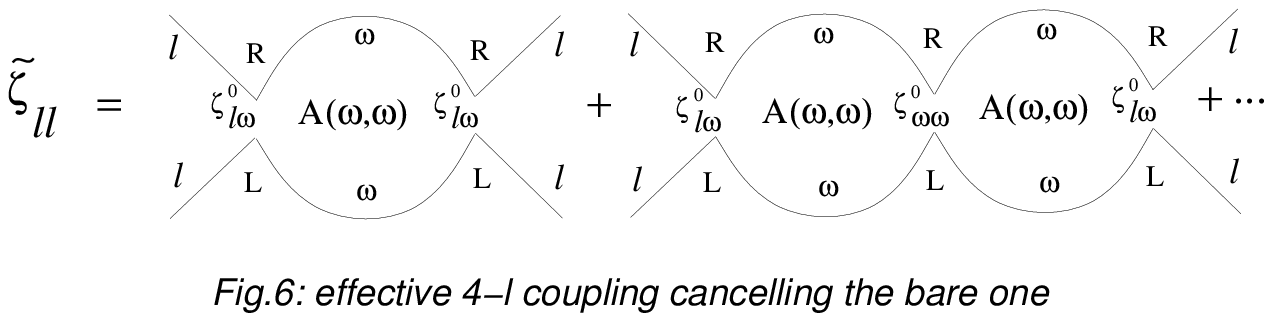}
\figskip
Owing to the four similar series yielding the same result which can be built
from the four $\ell\ell\omega\omega$ couplings of ${\cal L}_c$, we have
\be
\tilde\zeta_{\ell\ell} + \zeta_{\ell\ell}^0 = 0.
\ee
This is only to be taken as an indication that further cancellations are
expected than those in the Nambu-Jona-Lasinio approximation, to which we
will return for the rest of the paper.

\subsection{Summary: effective couplings and the mass eigenstates}

A demonstration at all orders of the renormalizability of our model is
beyond the scope of this work. We hope that the arguments given above have
convinced the reader that the perturbative expansion is much better behaved
that could be naively expected from the presence of 4-fermions couplings and
that the special type of Nambu-Jona-Lasinio mechanism invoked deserves
further investigation.

As announced, the ``see-saw'' mechanism exhibited at tree level has not been
altered by the
resummations above. The neutrino mass eigenstates are the Majorana $\chi$
and $\omega$ themselves, with respectively infinite and vanishing mass.
The charged lepton keeps its Dirac mass $m$, as contributions from 4-leptons
couplings can only be of the type of fig.~4 and vanish.

\section{The decoupling of the scalars}

When introducing a kinetic term for $\Delta$, the non-vanishing of the vacuum
expectation value $\la\Delta^0\ra$ contributes to the mass of the $Z$ and $W$
gauge bosons (see for example \cite{ChengLi}). If $v$ if the vacuum expectation
value of the hadronic Higgs boson (see \cite{BellonMachet1}),
we impose its role to be dominant, which yields the necessary condition
\be
                          \rho <\!\! < v
\ee
consistent with an electroweak nature for $\rho$ (see fig.~1), while that of
$v$
lies {\sl a priori} outside the realm of these interactions.

We shift in the usual way the neutral scalar field according to
\be
                          \Delta^0 = \rho + \delta^0.
\ee
{}From the expression of ${\cal L}_c$, we see that the non-vanishing of $\rho$
yields an infinite mass for $\delta^0$, $\Delta^+$, $\Delta^{++}$ and their
conjugates.
All those fields will decouple and are undetectable. In the limit $\beta\rar
0$, the longitudinal degrees of freedom of the massive gauge bosons reduce
to the
hadronic triplet $\vec\Phi$ \cite {BellonMachet1,GelminiRoncadelli}.

\section{The ${\bf V-A}$ theory}

The (massless) $\chi$ neutrino has the standard weak $V-A$ couplings and we can
identify it with the observed neutrino.

The $\omega$ neutrino is infinitely massive and will never be produced as
asymptotic state; we however expect renormalization effects
through $\omega$ loops \cite {ApplequistCarazzone}.
They drastically affect the neutral weak couplings of the
leptons, in a way that rebuilds their ``standard'' $V-A$ structure. This
result,
non-trivial if one remembers that the original coupling is purely vectorial, is
shown below.

To the bare (purely vectorial) couplings of $W_\mu^3$
we must add the following diagrams
\figskip
\hskip 1truecm
\epsffile{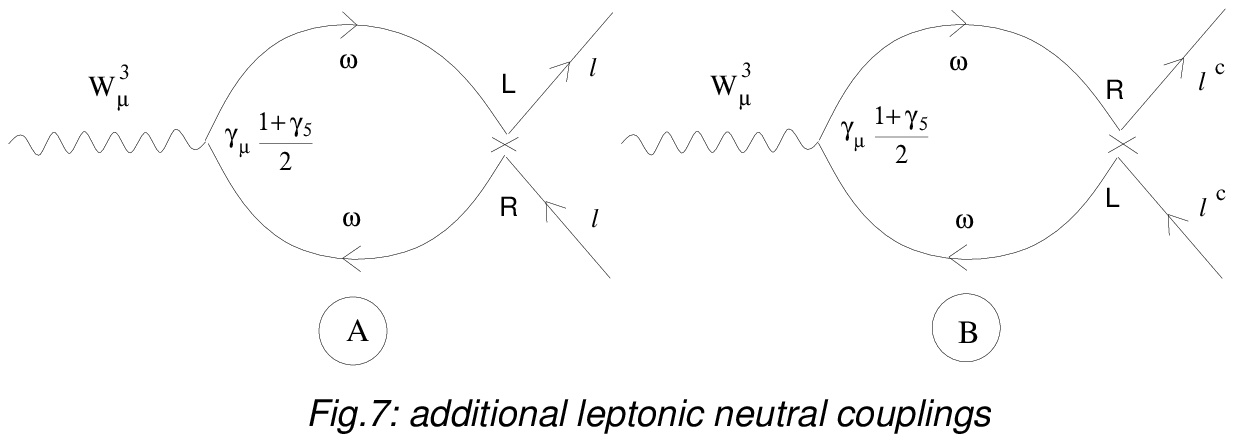}
\figskip
In fig.~7, the 4-fermions vertex is the effective $(\ell\ell\omega\omega)$
coupling $\zeta_{\ell\omega}$ obtained in 4.2, vanishing like $\beta^2$.
  Because of this dependence in $\beta$, diagrams similar to fig.~7 but
with $\omega$ replaced with $\chi$ vanish. Those involving
$\omega$ do not because the $\omega$ loop behaves like $\beta^{-2}$. Using the
$\ol{MS}$ renormalization scheme \cite{'t Hooft}, we indeed find, when
$M\rar\infty$:
\be\ba{rcl}
A &=&\phantom{-}{i\over 16\pi^2}\,
{\displaystyle\frac{M^2\ln M^2 +\cdots}{A_{\ell\omega} (q^2,m,M)}}
\; {g\over 2}\; W_\mu^3\, \ol\ell\gamma^\mu {1+\gamma_5\over 2}\ell,\cr
\noalign{\vskip 2mm}
B &=&-{i\over 16\pi^2}\,
{\displaystyle\frac{M^2\ln M^2 +\cdots}{A_{\ell\omega} (q^2,m,M)}}
\; {g\over 2}\; W_\mu^3\,  \ol{\ell^c}\,\gamma^\mu{1-\gamma_5\over 2}\ell^c,\cr
A_{\ell \omega}(q^2,m,M) &=& +{i\over 16\pi^2}\, 2M^2\ln M^2 +\cdots.
\ea\ee
Using
\be
\ol{\ell^c}\,\gamma_\mu{1-\gamma_5\over 2}\, \ell^c =
                       -\ol\ell\,\gamma_\mu {1+\gamma_5\over 2}\, \ell,
\ee
we see that fig.~7 contributes to a coupling
\be
+{g\over 2}\; W_\mu^3\, \ol\ell\,\gamma^\mu {1+\gamma_5\over 2}\, \ell,
\ee
such that  $W_\mu^3$ couples finally to
\be
-{g\over 2}\;\ol\ell\,\gamma^\mu\, \ell +{g\over 2}\;\ol\ell\,
\gamma^\mu {1+\gamma_5\over 2}\, \ell
            = -{g\over 2}\;\ol\ell\,\gamma^\mu {1-\gamma_5\over 2}\, \ell\, ,
\ee
which is the ``standard'' $V-A$ coupling.

Similarly, the coupling of $B_\mu$ to the charged leptons gets affected
by the same mechanism.
However, it couples to $\omega$ with a `$-$' sign with respect to the
corresponding coupling of $W_\mu^3$ because the hypercharge of $\omega$ is
minus the corresponding value of ${\Bbb T}^3$. This yields a final coupling
of $B_\mu$ to leptons
\be
-{g'\over 2}\; \bar\ell\,\gamma_\mu\,\ell -{g'\over 2}\; \bar\ell\,\gamma_\mu
{1+\gamma_5\over 2}\, \ell = -{g'\over 2}\; (2\; \bar\ell\,\gamma_\mu
{1+\gamma_5 \over 2}\, \ell + \bar\ell\,\gamma_\mu
{1-\gamma_5\over 2}\, \ell)\, ,
\ee
which is the usual coupling of the Standard Model, corresponding to an
hypercharge $-2$ for right handed leptons and $-1$ for left-handed ones.

The charged couplings do not get modified with respect to their bare values
since the diagram equivalent to fig.~7 depicted in fig.~8 involving an
infinitely massive
$\omega$ behaves like $\beta^2 M\ln M$ and so vanishes with $\beta$.
\figskip
\hskip 2.5cm
\epsffile{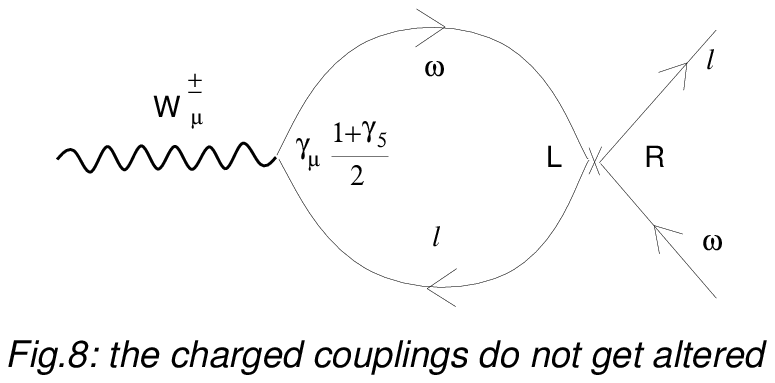}

\figskip

Ours is thus presently experimentally undistinguishable from the Standard
Model.

\section{Conclusion}

This work constitutes the second step in our effort to give another status
to the queer features of electroweak interactions than that of simple
postulates \cite{GlashowSalamWeinberg}; this goes in particular through cutting
the existing link between the hadronic and leptonic sectors.

The results that have been obtained are not
trivial: they yield a natural origin for the ``see-saw'' mechanism,
predict a massless observed neutrino and provide a link between the
unobservability of its partner and the $V-A$ structure of the weak currents.

{}From a more conceptual point of view, difficulties linked with Weyl fermions
are serious enough to welcome the disappearance of anomalies as a neat
progress.
We also propose a yet unexplored realization of the Nambu-Jona-Lasinio
mechanism
\cite{NambuJonaLasinio}, where both the bare fermion mass and the bare
4-fermions coupling are infinite.

It is clear that forthcoming efforts should concern renormalizability, since
the argumentation at the one-loop level is only a hint that our model behaves
nicely in the ultraviolet regime.

The first step of this program is performed in
\cite{Machet} for the abelian case and in \cite{BellonMachet1} for the
non-abelian Standard Model: we show there how the (usual) Higgs boson and the
quarks become unobservable, how the anomaly cancel and how the partners of the
Higgs in the scalar multiplet are in exact correspondence with observed
pseudoscalar mesons \cite{BellonMachet2}.

We clearly do not predict heaps of
new particles, or new scales of interactions and rather favour
economy as an attractive feature for a model.

\medskip {\em {\bf Acknowledgements}: it is a pleasure to thank P. Fayet for
comments and advice.}
\newpage\null
\listoffigures
\bigskip
\begin{em}
Fig. 1: contributions to the vacuum expectation value of the $\Delta^0$
boson.\l
Fig. 2: effective 4$\omega$ coupling.\l
Fig. 3: the effective $\omega$ mass.\l
Fig. 4: not a contribution to the $\omega$ mass.\l
Fig. 5: an effective $\ell\ell\omega\omega$ coupling.\l
Fig. 6: effective $4\ell$ coupling cancelling the bare one.\l
Fig. 7: additional leptonic neutral couplings.\l
Fig. 8: the charged couplings do not get altered.\l

\newpage\null

\end{em}
\end{document}